\documentclass[oldversion,print]{aa}

\usepackage{natbib}

\usepackage{graphicx}

\usepackage{txfonts}

\begin{document}

\title{The plasma emission model of RBS1774}

   \subtitle{}

  \author{Chkheidze, N
          }

   \institute{Center for theoretical Astrophysics, ITP, Ilia State
University, 0162, Tbilisi, Georgia\\
              \email{nino.chkheidze@iliauni.edu.ge}}

  \date{Received October 27, 2010; accepted December 15, 2010}


\abstract{In the present paper we construct a self-consistent
theory, interpreting the observational properties of RBS1774. It is
well known that the distribution function of relativistic particles
is one-dimensional at the pulsar surface. However, cyclotron
instability causes an appearance of transverse momenta of
relativistic electrons, which as a result, start to radiate in the
synchrotron regime. We study the process of the quasi-linear
diffusion developed by means of the cyclotron instability on the
basis of the Vlasov's kinetic equation. This mechanism provides
generation of measured optical and X-ray emission on the light
cylinder lengthscales. A different approach of the synchrotron
theory is considered, giving the spectral energy distribution that
is in a good agreement with the XMM-Newton observational data. We
also provide the possible explanation of the spectral feature at
$0.7$keV, in the framework of the model. }

\maketitle


\section{Introduction}

RBS1774 (1RXS J214303.7+065419) has been the most recent XDIN (X–ray
dim isolated neutron star) to be found \citep{zam01}. Its X-ray
spectrum is well reproduced by an absorbed blackbody with a
temperature $kT\sim100$eV and with a total column density of
$n_{H}\sim3\cdot10^{20}cm^{-2}$. Application of more sophisticated,
and physically motivated models for the surface emission
(atmospheric models) result in worse agreement with the data
\citep{zane05}. According to \citet{sch09}, a fit to the X-ray
spectra extracted from RGS spectrographs onboard XMM-Newton yields
that the best result is obtained when the two-temperature blackbody
model is used. But the same model applied to the X-ray spectra
extracted from three EPIC detectors does not improve the fit
compared to the simple blackbody model. However, the formation of a
non-uniform distribution of the surface temperature is more likely
artificial and needs to be examined by convincing theory.

Alternatively, the observational properties of RBS1774 can be
explained in the framework of the plasma emission model first
developed by \citet{mach79} and \citet{lomi}. According to these
works, in the electron-positron plasma of a pulsar magnetosphere the
waves excited by the cyclotron resonance interact with particles,
leading to the appearance of pitch angles, which obviously causes
synchrotron radiation. We suppose that the X-ray emission from this
object is generated by the synchrotron mechanism. According to the
standard theory of the synchrotron emission \citep{bekefi,ginz81}
the typical synchrotron spectrum is a power-law, when the present
model suggests different spectral distribution. The main reason for
this is that we take into account the mechanism of creation of the
pitch angles, consequently restricting their values. Contrary to
this, in the standard theory of the synchrotron radiation, it is
assumed that along the line of sight the magnetic field is chaotic,
leading to the broad interval (from $0$ to $\pi$) of the pitch
angles. The present model gives successful fit for the observed
X-ray spectrum, when the originally excited cyclotron modes enter
the same domain as the measured optical emission of RBS1774. We
suppose that the observed spectral feature at $0.7$keV in the X-ray
spectrum of RBS1774 is caused by wave damping process developed near
the light cylinder due to the cyclotron instability.

In this paper, we describe the emission model (Sec. 2), derive
theoretical X-ray spectrum of RBS1774 and fit with XMM-Newton
observations (Sec. 3), explain the possible nature of the observed
spectral feature at $\sim0.7$keV (Sec.4), and discuss our results
(Sec. 5).

\section{Emission model}

\begin{figure}
   \centering
\includegraphics[width=5cm]{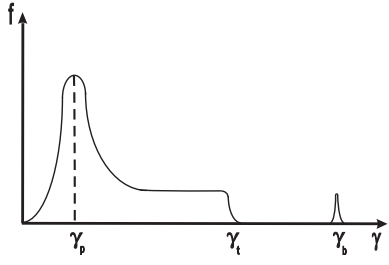}
      \caption{Distribution function of a
one-dimensional plasma in the pulsar magnetosphere. Left corresponds
to secondary particles, right to the primary beam.
              }
         \label{distr}
   \end{figure}

The distribution function of relativistic particles is one
dimensional at the pulsar surface, because any transverse momenta
($p_{\perp}$) of relativistic electrons are lost in a very short
time(\(\leq10^{-20}\)s) via synchrotron emission in very strong
magnetic fields. For typical pulsars the plasma consists of the
following components: the bulk of plasma with an average
Lorentz-factor $\gamma_{p}\simeq10^{2}$, a tail on the distribution
function with $\gamma_{t}\simeq10^{5}$, and the primary beam with
$\gamma_{b}\simeq10^{7}$ (see Fig.~\ref{distr}). However, plasma
with an anisotropic distribution function becomes unstable, which
can lead to a wave excitation in the pulsar magnetosphere. The
generation of waves is possible during the further motion of the
relativistic particles along the dipolar magnetic field lines if the
condition of cyclotron resonance is fulfilled \citep{kaz91b}:
\begin{equation}\label{1}
    \omega-k_{_{\|}}V_{_{\|}}-k_xu_x+\frac{\omega_{B}}{\gamma_{r}}=0,
\end{equation}
where $u_{x}=cV_{\varphi}\gamma_{r}/\rho\omega_{B}$ is the drift
velocity of the particles due to curvature of the field lines,
$\rho$ is the radius of curvature of the field lines and
$\omega_{B}=eB/mc$ is the cyclotron frequency. During the generation
of waves by resonant particles, one also has a simultaneous feedback
of these waves on the electrons  \citep{vvs}. This mechanism is
described by quasi-linear diffusion, leading to the diffusion of
particles as along as across the magnetic field lines. Therefore,
resonant electrons acquire transverse momenta (pitch angles) and, as
a result, start to radiate through the synchrotron mechanism.

The kinetic equation for the distribution function of the resonant
particles can be written as \citep{mach79,mach02}:
\begin{eqnarray}\label{2}
\frac{\partial\textit{f }^{0}}{\partial
    t}+\frac{\partial}{\partial
p_{\parallel}}\left\{(G_{\parallel}+F_{\parallel}+Q_{\parallel})\textit{f
}^{0}\right\}+\frac{1}{p_{\perp}}\frac{\partial}{\partial
p_{\perp}}\left\{p_{\perp}(G_{\perp}+F_{\perp})\textit{f
}^{0}\right\}=\nonumber
\\=\frac{1}{p_{\perp}}\frac{\partial}{\partial p_{\perp}}\left\{p_{\perp}\left(D_{\perp,\perp}\frac{\partial}{\partial p_{\perp}}+D_{\perp,\parallel}\frac{\partial}{\partial
p_{\parallel}}\right)\textit{f
}^{0}\left(\mathbf{p}\right)\right\}+\nonumber
 \\
+\frac{\partial}{\partial
p_{\parallel}}\left\{\left(D_{\parallel,\perp}\frac{\partial}{\partial
p_{\perp}}+D_{\parallel,\parallel}\frac{\partial}{\partial
p_{\parallel}}\right)\textit{f }^{0}\left(\mathbf{p}\right)\right\}.
\end{eqnarray}
where $G$ is the force responsible for conserving the adiabatic
invariant $p_{\perp}^{2}/B(r)=$const, $F$- is the radiation
deceleration force produced by synchrotron emission, and
$Q_{\parallel}$ is the reaction force of the curvature radiation.
They can be written in the form:
\begin{equation}\label{3}
    G_{\perp}=-\frac{mc^{2}}{\rho}\gamma_{r}\psi,\qquad \qquad G_{\parallel}=\frac{mc^{2}}{\rho}\gamma_{r}\psi^{2},
\end{equation}
\begin{equation}\label{4}
    F_{\perp}=-\alpha_{s}\psi\left(1+\gamma_{r}^{2}\psi^{2}\right),\qquad
    F_{\parallel}=-\alpha_{s}\gamma_{r}^{2}\psi^{2},
\end{equation}
\begin{equation}\label{5}
    Q_{\parallel}=-\alpha_{c}\gamma_{r}^{4},
\end{equation}
where $\alpha_{s}=2e^{2}\omega_{B}^{2}/3c^{2}$,
$\alpha_{c}=2e^{2}/3\rho^{2}$ and $\psi\approx
p_{\perp}/p_{\parallel}\ll1$ is the pitch angle.

Now let us compare the transverse components of the forces $G$ and
$F$. If we  consider the case $\gamma \psi\gg1$ we will have :
\begin{equation}\label{6}
     \frac{G_{\perp}}{F_{\perp}}=\frac{3m^{3}c^{6}}{2e^{4}B_{s}^{2}}\frac{\gamma}{\rho}\left(\frac{r}{R_{s}}\right)^{6}\frac{1}{\gamma^{2}\psi^{2}}.
\end{equation}
where $B_{s}$ is the magnetic field at the star surface, $R_{s}$ is
the star radius and $r$ is the distance from the pulsar. For the
typical parameter values of pulsars $|G_{\perp}|\ll|F_{\perp}|$.
Then taking into account that
$\partial/\partial\psi\gg\partial/\partial\gamma$ the equation for
the diffusion across the magnetic field can be written in the form
\citep{ch10}
\begin{eqnarray} \label{7}
    \frac{\partial\textit{f }^{0}}{\partial
    t}+\frac{1}{p_{\perp}}\frac{\partial}{\partial p_{\perp}}\left(p_{\perp}
    F_{\perp}\textit{f }^{0}\right)=\frac{1}{p_{\perp}}\frac{\partial}{\partial p_{\perp}}\left(p_{\perp}
D_{\perp,\perp}\frac{\partial\textit{f }^{0}}{\partial
p_{\perp}}\right).
\end{eqnarray}
where
\begin{equation}\label{8}
    D_{\perp,\perp}=\frac{e^{2}}{8c}\delta|E_{k}|^{2}
\end{equation}
is the diffusion coefficient and $|E_{k}|^{2}$ is the density of
electric energy in the waves.

The transversal quasi-linear diffusion increases the pitch-angle,
whereas force $\textbf{F}$ resists this process, leading to the
stationary state ($\partial\textit{f}/\partial t=0$). Then the
solution of Eq. (7) is
\begin{equation}\label{9}
    \textit{f}_{\perp}=C exp\left(\int
    \frac{F_{\perp}}{D_{\perp,\perp}}dp_{\perp}\right)=Ce^{-\left(\frac{p_{\perp}}{p_{\perp_{0}}}\right)^{4}}.
\end{equation}
To evaluate $p_{\perp_{0}}$, we use the quantity
\begin{equation}\label{10}
    |E_{k}|^{2}\approx\frac{1}{2}\frac{mc^{2}n_{b}\gamma_{b}c}{\omega},
\end{equation}
where $\omega$  is the frequency of original waves, excited during
the cyclotron resonance and can be estimated from Eq. (1) as follows
$\omega\approx\omega_{B}/\delta\gamma_{r}$. Consequently, we will
get
\begin{equation}\label{11}
     p_{\perp_{0}}\approx\frac{\pi^{1/2}}{B\gamma_{p}^{2}}\left(\frac{3m^{9}c^{11}\gamma_{b}^{5}}{32e^{6}P^{3}}\right)^{1/4}.
\end{equation}
The mean value of the pitch-angle $\psi_0\approx
p_{\perp_{0}}/p_{\parallel}\simeq10^{-3}$ (i. e. the assumption
$\gamma_{r}\psi_{0}\gg1$ done at the beginning of our computations
proves to be true). As a result of the appearance of the pitch
angles, the synchrotron emission is generated.

\section{X-ray spectrum}

Let us consider the synchrotron emission of the set of electrons.
The number of emitting particles in the elementary $dV$ volume is
$p_{\perp}\textit{f }d p_{\perp}dp_{\parallel}dVd\Omega_{\tau}$,
with momenta from the intervals $[p_{\perp}, p_{\perp}+dp_{\perp}]$
and $[p_{\parallel},p_{\parallel}+dp_{\parallel}]$, and with the
velocities that lie inside the solid angle $d\Omega_{\tau}$ near the
direction of $\vec{\tau}$. If we write the parallel distribution
function of the emitting particles as $\int
p_{\perp}\textit{f}dp_{\perp}\equiv\textit{f}_{\parallel}(p_{\parallel})$,
then the emission flux of the set of electrons will be
\citep{ginz81}
\begin{equation}\label{12}
    F_{\epsilon}=\int I_{e}\textit{f}_{\parallel}(p_{\parallel})dp_{\parallel}dVd\Omega_{\tau},
\end{equation}
where $I_{e}$ is the Stokes parameter, which is additive in this
case, as the observed synchrotron radiation wavelength $\lambda$ is
much less than the value of $n^{-1/3}$ - the average distance
between particles, where $n$ is the density of plasma component
electrons. The integral (\ref{12}) is easily reduced to (see
\citet{ginz81})
\begin{equation}\label{13}
    F_{\epsilon}\propto\int\textit{f}_{\parallel}(p_{\parallel})B\psi\frac{\epsilon}{\epsilon_{m}}\left[\int_{\epsilon/
    \epsilon_{m}}^{\infty}K_{5/3}(z)dz \right] dp_{\parallel}.
\end{equation}
Here $\epsilon_{m}\approx5\cdot10^{-12}B\psi\gamma^{2}$keV is the
photon energy of the maximum of synchrotron spectrum of a single
electron and $K_{5/3}(z)$ is a Macdonald function. After
substituting the mean value of the pitch-angle in the above
expression for $\epsilon_{m}$, we get
\begin{equation}\label{14}
    \epsilon_{m}\simeq5\cdot10^{-12}\frac{\pi^{1/2}}{\gamma_{p}^{2}}\left(\frac{3m^{5}c^{7}\gamma_{b}^{9}}{32e^{6}P^{3}}\right)^{1/4}
\end{equation}
For the primary beam electrons with the Lorentz factor
$\gamma_{b}\sim10^{7}$ the emitted photon energy
$\epsilon_{m}\sim0.1$keV comes in the energy domain of the observed
X-ray emission of RBS1774. Thus we suppose that the measured X-ray
spectrum is the result of the synchrotron emission of primary beam
electrons (the resonance occurs on the right slope of the
distribution function of beam electrons (see Fig.~\ref{distr})),
switched on as the result of acquirement of pitch angles by
particles during the quasi-linear stage of the cyclotron
instability.

To find the synchrotron flux in our case, we need to know the
one-dimensional distribution function of the emitting particles $
\textit{f}_{\parallel}$. Let us multiply both sides of Eq. (2) on
$p_{\perp}$ and integrate it over $p_{\perp}$. Using Eqs. (3), (4),
(5) and the following expressions for the diffusion coefficients
\citep{ch10}
\begin{eqnarray}\label{15}
    D_{\perp, \parallel}=D_{\parallel\perp} =-
    \frac{e^{2}}{8c}\psi|E_{k}|^{2},\\ \nonumber
    D_{\parallel,\parallel}=\frac{e^{2}}{8c}\psi^{2}\frac{1}{\delta}|E_{k}|^{2}.
\end{eqnarray}
And also taking into account that the distribution function vanishes
at the boundaries of integration, Eq. (2) reduces to
\begin{eqnarray}\label{16}
    \frac{\partial\textit{f}_{\parallel}}{\partial t}=\frac{\partial}{\partial
    p_{\parallel}}\left[\left({\frac{\alpha_{s}}{m^{2}c^{2}\pi^{1/2}}p_{\perp_{0}}^{2}}+\alpha_{c}\gamma^{4}-\frac{e^{2}}{4mc^{2}\gamma}|E_{k}|^{2}\right)\textit{f}_{\parallel}\right].
\end{eqnarray}
Let us estimate the contribution of different terms on the righthand
side of Eq. (16). The estimations show that the first term is much
bigger than two other terms. Consequently, for the primary-beam
electrons instead of Eq. (16), one gets
\begin{eqnarray}\label{17}
    \frac{\partial\textit{f}_{\parallel}}{\partial t}=\frac{\partial}{\partial
    p_{\parallel}}\left({\frac{\alpha_{s}}{m^{2}c^{2}\pi^{1/2}}p_{\perp_{0}}^{2}\textit{f}_{\parallel}}\right).
\end{eqnarray}

Considering the quasi-stationary case we find
\begin{eqnarray}\label{18}
    \textit{f}_{\parallel}\propto\frac{1}{p_{\parallel}^{1/2}|E_{k}|}.
\end{eqnarray}

For $\gamma\psi\ll10^{10}$, a magnetic field inhomogeneity does not
affect the process of wave excitation. The equation that describes
the cyclotron noise level, in this case, has the form \citep{lomi}
\begin{equation}\label{19}
    \frac{\partial|E_{k}|^{2}}{\partial
    t}=2\Gamma_{c}|E_{k}|^{2}\textit{f}_{\parallel},
\end{equation}
where
\begin{equation}\label{20}
   \Gamma_{c}=\frac{\pi^{2}e^{2}}{k_{\parallel}}\textit{f}_{\parallel}(p_{res}),
\end{equation}
is the growth rate of the instability. Here $k_{\parallel}$ can be
found from the resonance condition (1)
\begin{equation}\label{21}
     k_{\parallel_{res}}\approx\frac{\omega_{B}}{c\delta\gamma_{res}}.
\end{equation}
Combining Eqs. (17) and (19) one finds
\begin{equation}\label{22}
    \frac{\partial }{\partial t}\left\{\textit{f}_{\parallel}-\alpha\frac{\partial}{\partial
    p_{\parallel}}\left(\frac{|E_{k}|}{p_{\parallel}^{1/2}}\right)\right\}=0,
\end{equation}
\begin{equation}\label{23}
    \alpha=\left(\frac{4}{3}\frac{e^{2}}{\pi^{5}c^{5}}\frac{\omega_{B}^{6}\gamma_{p}^{3}}{\omega_{p}^{2}}\right)^{1/4},
\end{equation}
which reduces to
\begin{equation}\label{24}
    \left\{\textit{f}_{\parallel}-\alpha\frac{\partial}{\partial
    p_{\parallel}}\left(\frac{|E_{k}|}{p_{\parallel}^{1/2}}\right)\right\}=const.
\end{equation}

Taking into account that for the initial moment the major
contribution of the lefthand side of the Eq. (24) comes from
$\textit{f}_{\parallel_{0}}$, the corresponding expression writes as
\begin{equation}\label{25}
   \textit{f}_{\parallel}-\alpha\frac{\partial}{\partial
    p_{\parallel}}\left(\frac{|E_{k}|}{p_{\parallel}^{1/2}}\right)=\textit{f}_{\parallel_{0}}.
\end{equation}
The distribution function $\textit{f}$ is proportional to
$n\sim1/r^{3}$, then one should neglect $ \textit{f}_{\parallel}$ in
comparison with $\textit{f}_{\parallel_{0}}$. Consequently, the
above equation reduces to
\begin{equation}\label{26}
   \alpha\frac{\partial}{\partial
    p_{\parallel}}\left(\frac{|E_{k}|}{p_{\parallel}^{1/2}}\right)+\textit{f}_{\parallel_{0}}=0.
\end{equation}
As we can see the function $E_{k}(p_{\parallel})$ drastically
depends on the form of the initial distribution of the primary beam
electrons. Here we assume that the initial energy distribution in
the beam has a Gaussian shape
 \begin{equation}\label{27}
    \textit{f}_{b _{0}}=\frac{n_{b}}{\sqrt{\pi}\gamma_{T}}\textrm{ exp}\left[-\frac{\left(\gamma-\gamma_{b}\right)^{2}}{\gamma_{T}^{2}}\right],
\end{equation}
where $\gamma_{T}\simeq10$ - is the half width of the distribution
function and $n_{b}=B/Pce$ is the density of primary beam electrons,
equal to the Goldreich-Julian density \citep{go69}. Since
$\gamma_{T}\ll\gamma_{b}$, this distribution is very close to
$\delta$-function. Consequently, the electron distribution can be
taken as monoenergetic.

In this case for the energy density of the waves we get
\begin{equation}\label{28}
     |E_{k}|^{2}\propto p_{\parallel}.
\end{equation}
The effective value of the pitch angle depends on $|E_{k}|^{2}$ as
follows
\begin{equation}\label{29}
    \psi_{0}=\frac{1}{2\omega_{B}}\left(\frac{3m^{2}c^{3}}{p_{\parallel}^{3}}\frac{\omega_{p}^{2}}{\gamma_{p}^{3}}|E_{k}|^{2}\right)^{1/4}.
\end{equation}

According to our emission model, the observed radiation comes from a
region near the light cylinder radius, where the magnetic field
lines are practically straight and parallel to each other
\citep{os09}, therefore, electrons with $\psi\approx\psi_0$
efficiently emit in the observer's direction.

Using expression (18), (28) and (29), and replacing the integration
variable $p_{\parallel}$ by $x=\epsilon/\epsilon_{m}$, from Eq. (13)
we will get
\begin{equation}\label{30}
    F_{\epsilon}\propto\epsilon^{-0.3}\int x^{0.3}\left[\int_{x}^{\infty}K_{5/3}(z)dz \right] dx.
\end{equation}

The energy of the beam electrons vary in a small interval. In this
case the integral (30) can be approximately expressed by the
following function
\begin{equation}\label{31}
    F_{\epsilon}\propto\epsilon^{0.3} exp(-\epsilon/\epsilon_{m}).
\end{equation}

We performed a spectral analysis by fitting the model spectrum (Eq.
(31)) absorbed by cold interstellar matter, with the combined data
extracted of the three EPIC X-ray cameras of the XMM-Newton
Telescope. The resulting $\chi^{2}=1.63$ and the amount of
interstellar matter $n_{H}=(3.36\pm0.2)\times10^{20}cm^{-2}$, which
appears to be close to the total Galactic absorption in the source
direction ($n_{H}=5\cdot10^{20}cm^{-2}$ \citet{di90}). The spectral
feature at $\sim0.7$keV that is mostly described as an absorption
edge or line \citep{zane05,sch09} is also evident in our case from
inspection of Fig.~\ref{1}. We find that adding an absorption edge
at $0.7$keV improves the fit, leading to a reduced $\chi^{2}=1.50$.
The best-fitting energy of the edge is $E_{edge}=0.679$keV, and the
optical depth is $\tau_{edge}=0.20$ (see Fig.~\ref{2}). The fitting
results are listed in Table 1. We suppose that existence of the
absorption feature in spectra of RBS1774 is caused by wave damping
at photon energies $\sim0.7$keV, which takes place near the light
cylinder.

   \begin{figure}
   \centering
\includegraphics[width=8cm]{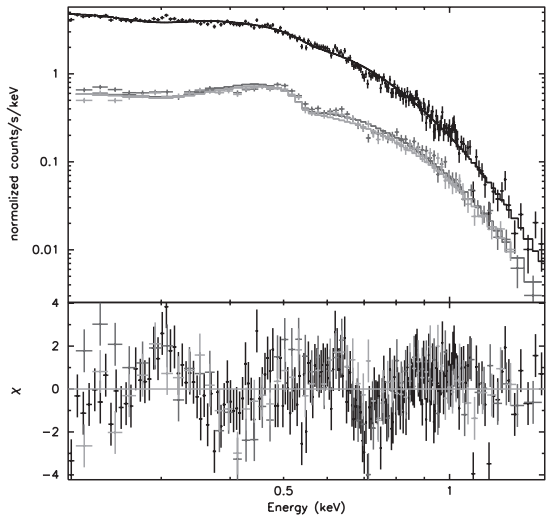}
      \caption{EPIC-pn and EPIC-MOS spectra of RBS1774, fitted with a model.
              }
         \label{1}
   \end{figure}

\section{Possible nature of the spectral feature}
During the farther motion in the pulsar magnetosphere, the X-ray
emission of RBS1774 that is generated on the light cylinder
lengthscales, might come in the cyclotron damping range
\citep{khe97}:
\begin{equation}\label{32}
     \omega-k_{_{\|}}V_{_{\|}}-k_xu_x-\frac{\omega_{B}}{\gamma_{r}}=0.
\end{equation}
The condition for the development of the cyclotron instability may
be easily derived for the small angles of propagation with respect
to the magnetic field. Representing the dispersion of the waves as
\begin{equation}\label{33}
    \omega=kc,
\end{equation}
and neglecting the drift term, the resonance condition (32) may then
be written as
\begin{equation}\label{34}
    \frac{1}{2\gamma_{r}^{2}}+\frac{\theta^{2}}{2}=\frac{\omega_{B}}{\omega \gamma_{r}},
\end{equation}
where $\theta\approx\psi$ is the angle between the wave vector and
the magnetic field. Taking into account that
$\psi_{0}^{2}\gg1/2\gamma_{b}^{2}$ one finds from Eq. (34) the
frequency of damped waves
\begin{equation}\label{35}
    \omega_{0}=\frac{2\omega_{B}}{\gamma_{r}\psi^{2}}.
\end{equation}
If we assume that the resonant particles are the primary beam
electrons, then the estimation shows that on the light cylinder
lengthscales $\epsilon_{0}=(h/2 \pi)\omega_{0}\simeq 0.7$keV.

   \begin{figure}
   \centering
\includegraphics[width=7.15cm]{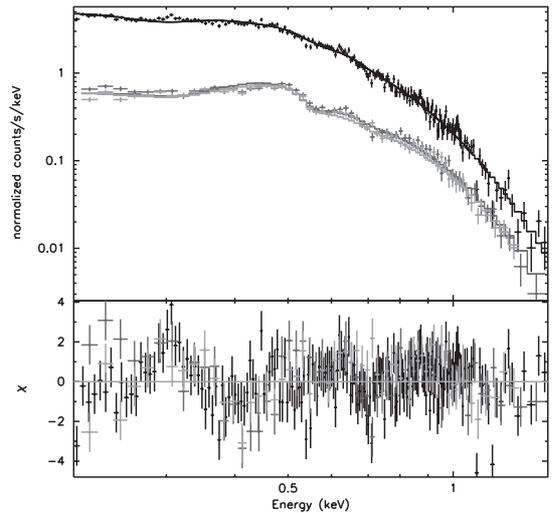}
      \caption{EPIC-pn and EPIC-MOS spectra of RBS1774 fitted with a model, including an absorption edge at $\sim0.7$keV.
              }
         \label{2}
   \end{figure}

\begin{table*}
 \caption{\label{table}The model parameters of RBS1774 for combined
fits to EPIC-pn and EPIC-MOS in the energy interval $0.2-1.5$
keV(The fitting results with a pure blackbody model absorbed by cold
interstellar matter are from \citet{sch09})} \centering
  \begin{tabular}{llrrrrlr}
  \hline \\
   Model    & $n_{H}$ & $\epsilon_{m}^{-1}$ & $kT_{bb}^{\infty}$ & $E_{edge/line}$ &  $\sigma_{line}$&$\tau_{edge/line}$ & $\chi^{2}$(dof)    \\
   & $(10^{20}cm^{-2})$ & (eV) & (eV) & (eV) & (eV) &
\\
\hline
\hline    \\                    
plasma & $3.36^{+0.20}_{-0.20}$ & $7.0\pm0.2$ &&&&& $1.63(311)$\\ \\
plasma*edge & $3.30^{+0.12}_{-0.12}$ & $6.9\pm0.1$ & & $679^{+13}_{-13}$ & & $0.20^{+0.03}_{-0.03}$ & $1.50(309)$\\ \\
bbody & $1.85^{+0.17}_{-0.17}$ &  & $103.5\pm0.8$ &  &  &   & $1.81(311)$\\
\\
bbody*gabs & $1.84^{+0.20}_{-0.17}$&  &  $105.1\pm0.9$ & $731^{+8}_{-13}$ & $27^{+16}_{-4}$ & $6.5^{+1.2}_{-1.0}$ & $1.50(308)$\\  
\hline
\end{tabular}

\end{table*}

\section{Discussion}
According to the generally accepted point of view, the X-ray
spectrum of RBS1774 is purely thermal and is best represented by a
Planckian shape. A fit with a pure blackbody component absorbed by
cold interstellar matter gives $\chi^{2}=1.81$ \citep{sch09}.
Including a Gaussian absorption line at $\sim0.7$keV (as the largest
discrepansies between model and data are around $0.7$keV) improves
the fit $\chi^{2}=1.50$ (parameters are listed in Tab.1). However,
the nature of this spectral feature is not fully clarified as yet.
The most likely interpretation is that it is due to proton cyclotron
resonance, which implies ultrastrong magnetic field of
$B_{cyc}\sim10^{14}$G \citep{zane05,re07}. Although, the required
strong magnetic field is inconstistent with timing measurements
giving $B_{dip}=3.2\cdot10^{19}\sqrt{P\dot{P}}\simeq2\cdot10^{13}$G
\citep{kap09}.

We are not about to reject the existing thermal emission models, but
in present paper we propose an alternative explanation of the
observed X-ray spectrum of RBS1774. It is supposed that the emission
of this source is generated by the synchrotron mechanism. The
distribution function of relativistic particles is one dimensional
at the pulsar surface, but plasma with an anisotropic distribution
function is unstable which can lead to wave excitation. The main
mechanism of wave generation in plasmas of the pulsar magnetosphere
is the cyclotron instability, which develops on the light cylinder
lengthscales. During the quasi-linear stage of the instability, a
diffusion of particles arises along and across the magnetic field
lines. Therefore, plasma particles acquire transverse momenta and,
as a result, the synchrotron mechanism is switched on. If the
resonant particles are the primary beam electrons with
$\gamma_{b}\simeq10^{7}$ their synchrotron emission enter the same
energy domain as the measured X-ray spectrum of RBS1774.

We construct a self-consistent theory interpreting the observations
of RBS1774. Differently from the standard theory of the synchrotron
emission \citep{ginz81}, which only provides a power-law spectrum
with the spectral index less than 1, our approach gives the
possibility to obtain different spectral energy distributions. In
the standard theory of the synchrotron emission, it is supposed that
the observed radiation is collected from a large spacial region in
various parts of which, the magnetic field is oriented randomly.
Thus, it is supposed that along the line of sight the magnetic field
directions  are chaotic and when finding emission flux, Eq. (13) is
averaged over all directions of the magnetic field (which means
integration over $\psi$ varying from $0$ to $\pi$). In our case the
emission comes from a region of the pulsar magnetosphere where the
magnetic field lines are practically straight and parallel to each
other. And differently from standard theory, we take into account
the mechanism of creation of the pitch angles. Thus we obtain a
certain distribution function of the emitting particles from their
perpendicular momenta (see Eq. (10)), which restricts the possible
values of the pitch angles. In the framework of the model we obtain
the following theoretical X-ray spectrum of RBS1774
$F_{\epsilon}\propto\epsilon^{0.3}exp(-\epsilon/\epsilon_{m})$ and
perform a spectral analysis by fitting data from the three EPIC
detectors simultaneously. The fit with a model spectrum absorbed by
cold interstellar matter yields $\chi^{2}=1.63$ (see parameters in
Tab.1).

During the farther motion in the pulsar magnetosphere, the X-ray
emission of RBS1774 comes in the cyclotron damping range (see Eq.
(32)). If we assume that damping happens on the left slope of the
distribution function of primary beam electrons (see
Fig.~\ref{distr}), then the photon energy of damped waves will be
$\epsilon_{0}=(h/2 \pi)\omega_{0}=(h/2
\pi)2\omega_{B}/\gamma_{b}\psi^{2}\simeq 0.7$keV. Taking into
account the shape of the distribution function of beam electrons, we
interpret the large residuals around $\sim0.7$keV (see Fig.~\ref{1})
as an absorption edge. Including an absorption edge improves the fit
leading to a reduced $\chi^{2}=1.50$. The best-fitting energy of the
edge is $E_{edge}=0.679$keV, and  the optical depth is
$\tau_{edge}=0.20$ (see Tab.1). However, adding an absorption edge
to the model spectrum does not produce a statistically significant
improvement of the fitting. According to \citet{sch09} if one uses
the RGS X-ray spectra of RBS1774 in place of EPIC spectra, the
resulting $\chi^{2}$ is changed just marginally when a Gaussian
absorption line is included at $\sim0.7$keV. Thus, we conclude that
the nature of the feature at $0.7$keV is uncertain and might be
related to calibration uncertainties of the CCDs and the RGS at
those very soft X-ray energies. The same can be told about a feature
at $\sim0.3$keV (the large residuals around $0.3$keV are evident
from inspection of Fig.~\ref{1} and ~\ref{2}). A feature of possible
similar nature was detected in EPIC-pn spectra of the much brighter
prototypical object RXJ1856.4-3754 and classified as remaining
calibration problem by \citet{hab07}. Consequently, more data are
necessary to finally prove or disprove the existence of those
features.

The frequency of the original waves, excited during the cyclotron
resonance can be estimated from Eq. (1) as follows
$\nu\approx2\pi\omega_{B}/\delta\gamma_{b}\sim10^{14}$Hz. As we can
see the frequency of cyclotron modes comes in the same domain as the
measured optical emission of RBS1774 \citep{zane08,sch09}.

\section*{Acknowledgments}

The author is grateful to George Machabeli for valuable discussions
and Axel Schwope for providing the X-ray data.

\end{document}